\documentstyle[aps,prb,multicol,epsf,amstex,verbatim]{revtex}

\begin{document}

\draft

\widetext

\title{Direct experimental verification of applicability of single-site 
model for angle integrated photoemission of small $T_{\rm K}$ 
concentrated Ce compounds}

\author{H.-D. Kim,$^{1}$ J. W. Allen,$^1$ J.-H. Park,$^2$ A. Sekiyama,$^3$
A. Yamasaki,$^3$ K. Kadono,$^3$ S. Suga,$^3$ Y. Saitoh,$^4$ T. Muro,$^5$
E. J. Freeman,$^6$ N. A. Frederick,$^6$ and M. B. Maple$^6$}

\address{$^1$Randall Laboratory of Physics, University of Michigan,
Ann Arbor, MI 48109, USA}
\address{$^2$Department of Physics, Pohang University of Science and
Technology, Pohang 790-784, Korea}
\address{$^3$Department of Material Physics, Osaka University, 
1-3 Machikaneyama, Toyonaka, Osaka 560-8531, Japan}
\address{$^4$Department of Synchrotron Radiation Research, Japan
Atomic Energy Research Institute, SPring-8,\\
Sayo, Hyogo 679-5143, Japan}
\address{$^5$Japan Synchrotron Radiation Research Institute,
SPring-8, Sayo, Hyogo 679-5143, Japan}
\address{$^6$Department of Physics, University of California at 
San Diego, La Jolla, CA 92093, USA}

\maketitle

\begin{abstract}
Bulk-sensitive high-resolution Ce 4f spectra have been obtained
from 3d $\rightarrow$ 4f resonance photoemission measurements on
La$_{1-x}$Ce$_x$Al$_2$ and La$_{1-x}$Ce$_x$Ru$_2$ for $x = 0.0,
0.04, 1.0$. The 4f spectra of low-Kondo-temperature ($T_{\rm K}$)
(La,Ce)Al$_2$ are essentially identical except for a slight
increase of the Kondo peak with $x$, which is consistent with a
known increase of $T_{\rm K}$ with $x$. In contrast, the 4f
spectra of high-$T_{\rm K}$ (La,Ce)Ru$_2$ show a Kondo-like peak
and also a 0.5~eV structure which increases strongly with $x$. The
resonance photon-energy dependences of the two contributions are
different and the origin of the 0.5~eV structure is still
uncertain.
\end{abstract}

\pacs{{\em Key words:} Ce compounds, photoemission}

\begin{multicols}{2}

\narrowtext

The single-impurity Anderson model (SIAM) has been used with great
success to describe angle integrated 4f photoemission spectra
of dense-Kondo Ce compounds, both the low energy Kondo features
and the high energy ionization excitations, after
proper subtraction of a surface spectrum \cite{siam}, 
in spite of claims to the contrary \cite{arko}. Theory
support for such claims might come from Nozi\`eres
`exhaustion' idea, that in the simple Kondo lattice
there are not enough conduction electrons to screen out the
magnetic moments as in the SIAM, essentially because only 
electrons located within the Kondo temperature ($T_{\rm K}$) 
of the Fermi level ($E_{\rm F}$) can 
take part in the Kondo screening \cite{nozieres}. The exhaustion
problem is thus most serious in the lowest-$T_{\rm K}$ systems, 
and would imply a decreased $T_{\rm K}$ in the lattice
system relative to the impurity system. In this work, we have made
a direct experimental comparison of 4f spectra of concentrated and
dilute Ce compounds by performing bulk-sensitive high-resolution Ce
3d $\rightarrow$ 4f resonant photoemission spectroscopy (RPES) of
La$_{1-x}$Ce$_x$Al$_2$ and La$_{1-x}$Ce$_x$Ru$_2$, for which the 
$T_{\rm K}$'s are 5~K and larger than 1000~K, respectively, in the
concentrated limit. They have the same cubic Laves structure in
which the number of nearest Ce neighbors of a Ce ion is only four.
For $x = 0.04$, 85\% of the Ce ions are completely 
isolated from other Ce ions if next-nearest Ce-Ce 
couplings are neglected, so that the 4f spectrum is 
dominated by the single-site contribution.

Polycrystalline samples of La$_{1-x}$Ce$_x$Al$_2$ and
La$_{1-x}$Ce$_x$Ru$_2$ ($x = 0.0, 0.04, 1.0$) were prepared by arc
melting and characterized by x-ray diffraction.  Some elemental Ru
impurity was found in the dilute (La,Ce)Ru$_2$
sample. Although the Ru impurity contribution should be removed 
by the RPES extraction of the Ce 4f spectrum, 
nonetheless the impurity renders our results for the 
(La,Ce)Ru$_2$ system to be tentative.  
Ce 3d $\rightarrow$ 4f RPES was performed at beam line BL25SU of
SPring-8 \cite{sekiyama}. 
All samples were fractured to reveal clean surfaces in a vacuum of 
$2\times10^{-10}$~Torr.
The overall energy resolution 
was about 100~meV and the sample temperature was kept at 20~K. 

Figure 1 shows a comparison of the Ce 4f spectra of concentrated
and dilute (La,Ce)Al$_2$ obtained from the RPES spectra 
by a usual method \cite{siam}.
The sharp two-peak structure
near $E_{\rm F}$ is a generic feature of low-$T_{\rm K}$ Ce
compounds, consisting of the tail of a Kondo resonance (KR) and
its spin-orbit replica (SOR) near 0.25~eV.  Other structures 
at higher binding
energy are the material-specific ionization features of the
hybridization matrix element $[V(\epsilon)]^2$ between Ce 4f and
conduction electrons, within the SIAM description. 
There is perfect agreement of the two spectra 
except for an overall reduction of
the KR and its SOR in the dilute system.  The known 
reduc-
\linebreak
\begin{figure}
\epsfxsize=85mm
\centerline{\epsffile{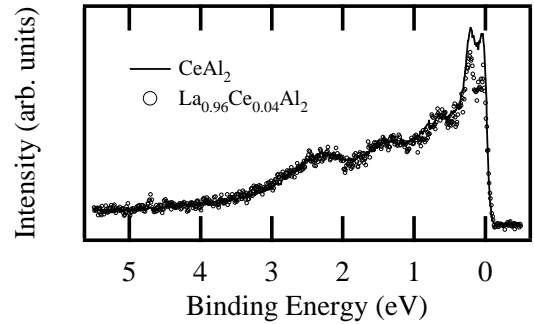}}
\caption{Ce 4f spectra of CeAl$_2$ and
La$_{0.96}$Ce$_{0.04}$Al$_2$ obtained by subtracting an
off-resonance spectrum from an on-resonance one at the Ce
3d $\rightarrow$ 4f edge.}
\end{figure}

\begin{figure}
\epsfxsize=85mm
\centerline{\epsffile{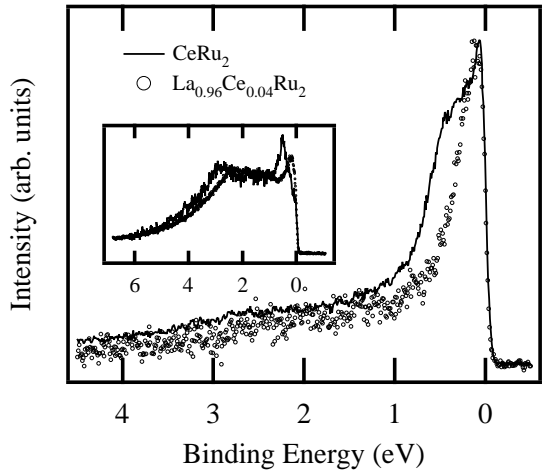}}
\caption{The same as in Fig. 1 except for (La,Ce)Ru$_2$. Inset shows
off-resonance spectra.}
\end{figure}

\noindent
tion of $T_{\rm K}$ with decreasing Ce concentration 
\cite{bredl}, due to volume expansion, is opposite 
to expectations from Nozi\`eres `exhaustion' idea 
but easily explains the small spectral change within a SIAM 
description, including the stronger intensity reduction 
of the KR relative to its SOR.  
The results are a direct experimental confirmation 
of the applicability of a single-site 
description of the angle integrated Ce 4f spectrum 
in this low-$T_{\rm K}$ system, contrary to objections 
\cite{arko} that have been made.  

The Ce 4f spectra of concentrated and dilute (La,Ce)Ru$_2$ are
presented in Fig.~2.  In great contrast to the (La,Ce)Al$_2$ 
system, the two spectra differ greatly, especially by a 
0.5~eV structure in the concentrated system.  
At present the origin of the 0.5~eV structure is unclear.  
The difference might well imply the need to go beyond a 
single-site model in the high-$T_{\rm K}$ system due to strong 
intersite couplings arising from large Ce 4f-Ru 4d hybridization 
\cite{sekiyama}.  But this conclusion may be premature. 
As shown in the inset, the off-resonance spectrum has a strong 
$x$-dependence, that a peaky 0.5~eV structure in the concentrated
system is shifted to $E_{\rm F}$ for the dilute system.  
Our calculation (not shown) of the 4f spectrum using the
Gunnarsson-Sch\"onhammer method \cite{gs} 
and taking $[V(\epsilon)]^2$ to be the off-resonance 
spectrum, does show the 0.5~eV structure in the concentrated system, 
though with an intensity that is rather smaller than in the experimental 
spectrum.  The fact that the 0.5~eV structure in CeRu$_2$ 
is found to become much stronger when the spectrum is obtained 
at a slightly higher photon energy motivates yet another explanation, 
that it arises from an incoherent Auger transition, as clearly 
observed in CeFe$_2$ \cite{cho}.  
Further experimental investigation is now in progress.

\vspace*{4mm}
This work was supported by the U.S. NSF at the
University of Michigan (No. DMR-99-71611), by the
U.S. DoE at the University of Michigan (No. DE-FG-02-90ER45416),
and by the U.S. NSF at the
University of California at San Diego (No. DMR-97-05454).

\end{multicols}


\begin{thebibliography}{0}

\bibitem{siam}J. W. Allen, {\it et al.}, Adv. Phys. 35 (1986) 275; 
D. Malterre, M. Grioni, and Y. Baer, Adv. Phys. 45 (1996) 299; 
J. W. Allen , {\it et al.}, J. Appl. Phys. 87 (2000) 6088;
A. Sekiyama, {\it et al.}, J. Phys. Soc. Jpn. 69 (2000) 2771.
\bibitem{arko}A. J. Arko, {\it et al.}, in {\it Handbook on the Physics
and Chemistry of Rare Earths}, Vol. 26, edited by K. A. Gschneidner 
and L. Eyring (Elsevier Science, B.V., 1999) pp. 265-382.
\bibitem{nozieres}Ph. Nozi\`eres, Eur. Phys. J. B 6 (1998) 447.
\bibitem{sekiyama}A. Sekiyama, {\it et al.}, Nature 403 (2000) 396.
\bibitem{bredl}C. D. Bredl, F. Steglich, and K. D. Schotte, 
Z. Physik B 29 (1978) 327.
\bibitem{gs}O. Gunnarsson and K. Sch\"onhammer, Phys. Rev. B 31 
(1985) 4815.
\bibitem{cho}E.-J. Cho and S.-J. Oh, (private communication).

\end{thebibliography}
\end{document}